# Interstitial Carbon in bcc HfNbTiVZr high entropy alloy from first principles

Luis Casillas-Trujillo[1], Ulf Jansson[2], Martin Sahlberg[2], Gustav Ek[2], Magnus M. Nygård[3], Magnus H. Sørby[3], Bjørn C. Hauback[3], Igor Abrikosov[1,4], Björn Alling[1]

[1]Department of Physics, Chemistry and Biology (IFM), Linköping University, 58183 Linköping, Sweden.

[2]Department of Chemistry-Ångström, Uppsala University, 75121 Uppsala, Sweden

[3] Department for Neutron Materials Characterization, Institute for Energy Technology, NO-2027 Kjeller, Norway

[4] Materials Modeling and Development Laboratory, National University of Science and Technology 'MISIS', Moscow, 119049 Russia

## Abstract

The remarkable mechanical properties of high entropy alloys can be further improved by interstitial alloying. In this work we employ density functional theory calculations to study the solution energies of dilute carbon interstitial atoms in tetrahedral and octahedral sites in bcc HfNbTiVZr. Our results indicate that carbon interstitials in tetrahedral sites are unstable, and the preferred octahedral sites present a large spread in the energy of solution. The inclusion of carbon interstitials induces large structural relaxations with long-range effects. The effect of local chemical environment on the energy of solution is investigated by performing a local cluster expansion including studies of its correlation with the carbon atomic Voronoi volume. However, the spread in solution energetics can not be explained with a local environment analysis only pointing towards a complex, long-range influence of interstitial carbon in this alloy.

1. Introduction

In recent years high entropy alloys (HEA) have gathered a vast interest due to their mechanical properties. In particular, HEA consisting of refractory elements have possible applications as high performance materials at elevated temperatures. Near equimolar WTaMoNb and WTaMoNbV have displayed better mechanical properties at elevated temperatures than conventional Ni based superalloys [1]. Besides mechanical properties, HEA have been investigated as highly tailorable hydrogen storage materials [2,3]. HfNbTiVZr has attracted special attention for its excellent hydrogen storage capacity [4], with experimental data suggesting that hydrogen can occupy both tetrahedral and octahedral positions in the face centered structured (fcc) phase [5,6]. Furthermore, $Hf_{21}Nb_{25}Ti_{15}V_{15}Zr_{24}$ has also been reported to display superconductivity [7]. HfNbTiVZr is believed to be a single-phase HEA with a body centered cubic (bcc) structure thermodynamically stable above 800 °C. At lower temperatures the thermodynamically stable phase is a mixture of bcc and hexagonal closed packed (hcp) alloys and a Laves phase [8]. However, metastable solid solution samples have been obtained at room temperature [9,10].

The properties of refractory HEAs can be further improved by interstitial solution of other elements in the structure. For example, the incorporation of carbon atoms into HEA has been studied experimentally [11-23] displaying an increase in yield and ultimate strengths. To understand the change in mechanical properties from the incorporation of carbon, an atomic scale understanding is needed. As such, *ab initio* based studies are a powerful tool and a key piece to develop strategies to fine-tune mechanical properties. Recently, Ikeda *et al.* [24] have studied the impact of interstitial carbon in fcc structured CrMnFeCoNi using first principles calculations. Carbon atoms prefer to occupy octahedral interstitial sites in bcc metals such as iron [25,26], Nb and V. The small size of the octahedral interstice produces a low maximum solid solubility of carbon in pure metals in group 4–6 of the Periodic Table, and therefore a low carbon solubility in solid bcc HEA would also be expected. However, carbon dissolution could be assisted by the different interstitial site sizes, produced by lattice distortions originating from the atomic size mismatch, in addition to the effect of local chemical environments. We address these research questions in the present work, and study the further distortions produced by the incorporation of carbon. Using density functional theory (DFT) calculations, we have investigated the effect of local lattice distortions and the effect of local chemical environments on the energy of solution of C interstitial atoms in tetrahedral and octahedral position in bcc HfNbTiVZr.

## 2. Methods

*2.1 Interstitial volumes*

There are 6 octahedral sites and 12 tetrahedral sites per unit cell in the bcc structure. Octahedral sites are located at the center of the faces and mid-points of the edges of the unit cell. Octahedral sites are six-fold coordinated, and the octahedron is distorted with two atoms closer to the center than the other four atoms. Tetrahedral sites are located between the center of the faces and mid points of each edge. Figure 1 shows the tetrahedral and octahedral sites in the bcc structure and the nearest neighbor polyhedra.

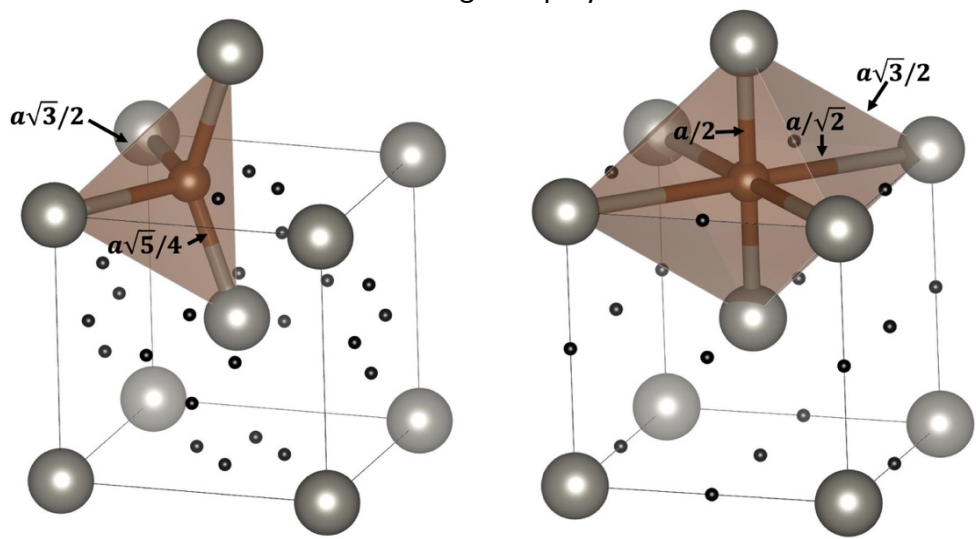

Figure 1. Tetrahedral and octahedral interstitial sites in the bcc structure.

Using a purely geometric approach, one could speculate that the best position to place an interstitial atom is the one with the most available free space. In a perfect bcc structure, the hard atom sphere model gives an atomic solute radius of $0.154 r_A$ for the octahedral site and $0.291 r_A$ for the tetrahedral site, where $r_A$ is the atomic radius of the constituent atoms. In the HEA case, the multiple atomic sizes of the alloy's components and the different local environments make the assessment of the void's spaces available for the dissolution of an interstitial atom into a non-trivial task. In this work, we estimate the available volume using the so-called radical Voronoi tessellation [25,26]. In this method atomic volumes take the form of irregular polyhedra with faces built from the planes which are perpendicular bisectors between a particle and its neighbors. The radical Voronoi tessellation is a generalization of the conventional Voronoi polyhedra procedure to examine multicomponent compounds. The resulting radical plane polyhedra depend on the relative sizes of the atoms, forming an assembly of polyhedra which completely fill space. The radical tessellation includes all points whose tangential distance to the surface of the sphere is smaller than the tangential distance to the other spherical surfaces.

In a HEA the nearest neighbor polyhedra around the interstices are distorted due to the relaxation effects caused by the presence of the multiple components [6,27,28]. We have tested different positions inside the nearest neighbor polyhedral to determine the most suitable candidate position to place an interstitial atom. The radical Voronoi volume for the centroid and for the midpoint of the vector connecting the centroid to the nearest neighbors is calculated, and the position with the largest volume is selected. The probe positions are shown in Figure 2. We have used the voro++ library for the Voronoi volume calculations [29].

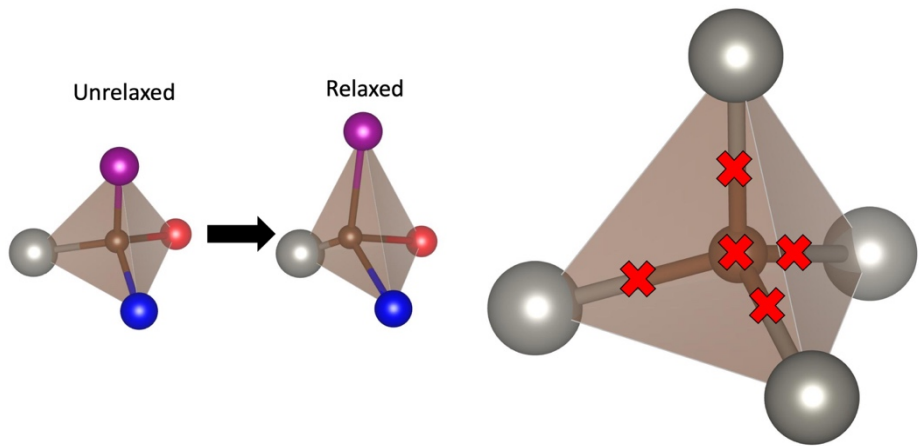

Figure 2. Schematic comparison of the perfect and distorted nearest neighbor polyhedra, and the positions probed to find the most suitable position to place an interstitial atom.

2.2 DFT calculations

The DFT calculations are performed using the projector augmented wave method [30] as implemented in the Vienna *Ab-initio* Simulation Package (VASP) [31,32] with the Perdew–Burke–Ernzenhof generalized gradient approximation [33] to model the exchange correlation effects. All calculations are performed using a 420 eV kinetic-energy cutoff, with a convergence criterion on forces of 0.01 eV/Å for structural relaxations. We use a 3x4x5 HEA supercell with 120 atoms, which allows for an equiatomic composition. The supercell is created using the special quasirandom structure method (SQS) [34] and contains 720 tetrahedral positions, and 360 octahedral positions. We have inserted one carbon interstitial into the relaxed HEA supercell for each of the obtained largest available volume positions of the tetrahedral and octahedral interstitial sites and further relaxed the volume, shape and atom positions of the system.

*2.3 Solution energies of interstitial atoms*

The solution energies $\Delta E_{sol}$ of C atoms are calculated as:

$$\Delta E_{sol} = E(alloy + C) - [E(alloy) + E(C)] \quad (1)$$

where $E(alloy + C))$ and $E(alloy)$ are the energies of the alloys with and without one C atom. $E(C)$ is the energy of graphite per atom. Due to the inadequate description of Van der Waals interactions of the GGA approximation, there is a large overestimation of the energy difference between graphite and diamond [25,35,36]. Experimentally it has been found that the enthalpy of graphite is 0.019 eV lower than the diamond enthalpy [37]. To obtain $E(C)$, we calculate the energy of diamond and shift it by applying the experimental difference value.

### 3. Results

*3.1 HfNbTiVZr without carbon*

The size mismatch and different element interactions give rise to considerable distortions in the relaxed system. The relaxation of the carbon free HEA HfNbTiVZr yields a lattice constant of 3.36 Å. Fazakas *et al.* reported a 3.40 Å lattice constant determined from theoretical calculations [9]. The experimental values of the lattice constant are 3.3659(2) Å [5], 3.36(5) [10], 3.385 [9] and 3.36466(9) Å [6]. In the relaxed structure, we find an average displacement from ideal bcc positions of 0.27 Å with a standard deviation of 0.13 Å. Table 1 presents the average of the displacements per atomic species. All atomic species display large displacements from their ideal bcc positions, but no particular atomic specie's behavior can be discerned due to the large standard deviations. In the relaxed structure, the different local chemical environments generate a distribution of interstitial volumes. In Figure 3 we have plotted the volumes for an interstitial carbon atom in tetrahedral and octahedral sites using our largest available volume approach with the radical Voronoi tessellation. We have obtained the pair distribution function of the relaxed HfNbTiVZr without carbon and find a very good agreement with the experimental neutron total scattering results [6] as shown in Figure 4. This validates our usage of a SQS structure modeling the HEA as a fully random alloy. There is a small shift of 0.09 Å in the nearest neighbor peak between calculations and the experimental data that might originate from lattice vibrations not included in the calculations or a small amount of short-range order in the experiments.

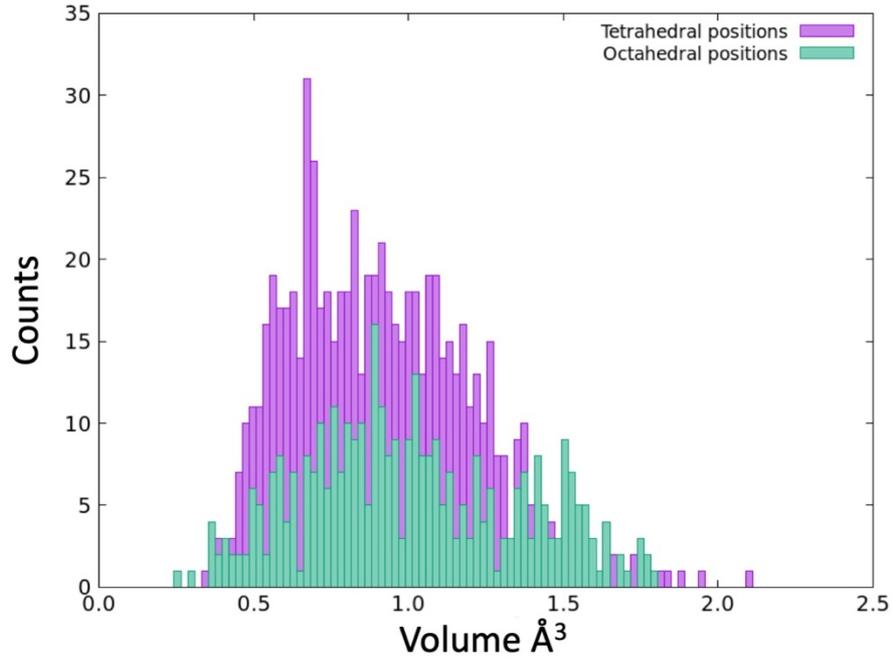

Figure 3. Radical Voronoi volume distribution of tetrahedral and octahedral sites in the relaxed HEA.

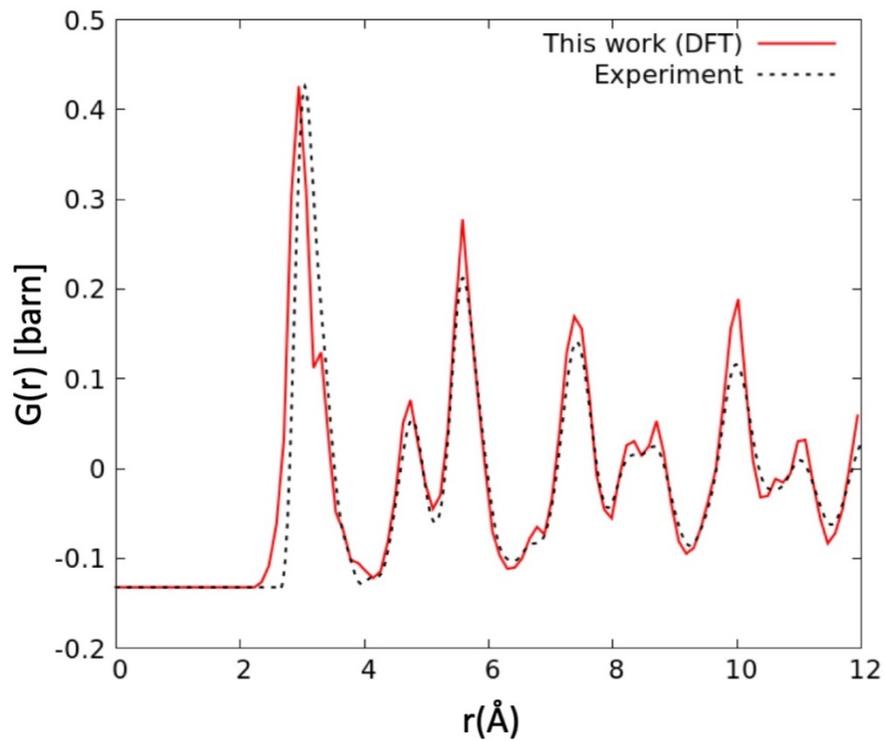

Figure 4. Pair distribution function of the DFT relaxed HfNbTiVZr and compared with the experimental neutron total scattering results in reference [6]. The function from DFT is scaled so that the baseline and the first peak are at the same level as in the experimental function.

Table 1. Average displacement per atomic species between ideal and relaxed HfNbTiVZr.

|    | Average (Å) |
|----|-------------|
| Hf | 0.25(15)    |
| Nb | 0.23(9)     |
| Ti | 0.30(12)    |
| V  | 0.31(10)    |
| Zr | 0.27(15)    |

*3.2 Energy of solution*

The initial energies of supercells with interstitial carbon placed in the carbon-free relaxed HEA, and the final energies after relaxing the supercells with the carbon atom for both tetrahedral and octahedral positions are shown in Figure 5. The *x* axis has been sorted with respect to interstitial void volume in ascending order in the plots. For the initial case, it is possible to discern between the behavior of the tetrahedral and octahedral cases. In particular, the octahedral sites display a large range of initial energies, with some octahedral sites having considerably higher and lower energies than the tetrahedral sites. For the initial case, there is a correlation between energy and volume, with larger volumes having lower energies for both the tetrahedral and octahedral positions. Figure 5b shows the final energy after relaxation. In this case, there is no clear correlation between the energy and the interstitial volume, and the energies of tetrahedral and octahedral positions are not clearly distinguishable. This is because during relaxation, the carbon atoms initially put on tetrahedral sites diffuse to octahedral sites or interstitial positions in between octahedral and tetrahedral sites. The diffusion of C could be corroborated by visualizing the atomic positions, but to assess the large number of cases we have employed the Steinhardt bond-orientational parameters [38] and used Pyscal [39] to calculate them (details in the supplementary materials). For the octahedral sites, the range in energy values has reduced, from ~20 eV in the initial case to ~2eV in the final case. This indicates that some initial positions were highly unfavorable, while the most favorable positions were not strongly affected by relaxations. All the carbon interstitials put at octahedral sites stay there. Due to this result, for the remainder of the paper we only show results for carbon in octahedral positions.

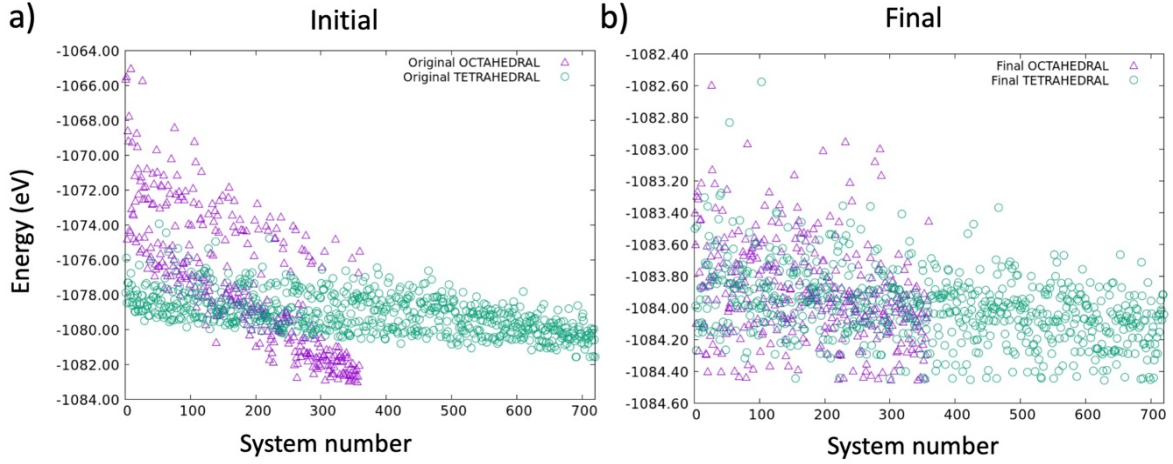

Figure 5. Total energies of supercells modeling HfNbTiVZr HEA with C interstitials in octahedral and tetrahedral sites. The initial energy corresponds to the total energy of supercells with interstitial carbon placed in the SQS modeling of a carbon-free relaxed HEA, while the final energy is calculated for the fully relaxed supercell with the carbon atom in the system. The $x$ axis shows each of the interstitial positions, which have been sorted in ascending radical Voronoi volume.

Figure 6 shows the distribution of $\Delta E_{sol}$ of carbon atoms at octahedral positions calculated according to Eq. 1, with an average value of -1.01 eV and standard deviation 0.31 eV. In comparison, we find that a carbon interstitial in an octahedral site in pure bcc Nb and pure bcc V has a solution energy of -0.67 eV and -0.74 eV, respectively, which may indicate that the multicomponent alloying favors C dissolution in the HEA. Moreover, the lowest solution energy in the HEA with a value of -1.56 eV (corresponding to a local environment comprising of V-Ti-Ti-Nb-Zr-Zr nearest neighbors) is significantly lower than the mean value. The highest energy corresponds to a Nb rich local environment with a solution energy of 0.29 eV. This is the only octahedral site with a positive energy of solution. Table 2 presents the energy of formation of the monocarbides (HfC, NbC, TiC, VC and ZrC) and the energy of solution of a carbon interstitial in an octahedral site for the pure bcc metals Nb and V, and hcp Hf, Ti and Zr. In the table we have also included the average energies for metal-type-$i$-rich ($M_i$-rich) local, environments around carbon interstitial atoms in the HEA. An environment is defined as $M_i$-rich when the $M_i$ element occupies half or more of the nearest neighbor sites. For the monocarbides, group 4 (Hf, Ti, Zr) compounds have a more negative formation energy suggesting stronger bonds than for those of group 5 (Nb, V). This trend is maintained in the energy of solution of the pure metals case, with the hierarchy of Nb and V reversed, whereas this is no longer the case for the HEA averages. Figure 7 shows $\Delta E_{sol}$ for $M_i$-rich local, environments around carbon interstitial atoms. Among the different $M_i$-rich environments, Ti-rich is the most favorable, with an average of -1.23 eV followed by V, Hf, Zr, with Nb being the least favorable. Therefore, it would be more likely to encounter carbon atoms close to Ti-rich local environments of the material. In Figure 7, we also have included the distribution for "HEA-like" local environments, meaning environments including all five metals with one repeated element type. For the HEA-like case, the lowest energy is -1.54 eV and highest energy -0.06 eV. A comparison with solution energies of the carbon interstitial in an octahedral site in pure bcc Nb and pure bcc V given above, shows that the values for the pure metals are 0.18 eV and 0.29 eV above the mean value of Nb-rich local and V-rich environments, respectively, found in our HEA. The amount of carbon that can be dissolved in the HEA will depend on the number of energetically favorable environments and carbon-carbon interactions. One can imagine that carbon atoms would first occupy the lowest energy sites, in a similar way as proposed for the creation of vacancies in metastable random alloys [40].

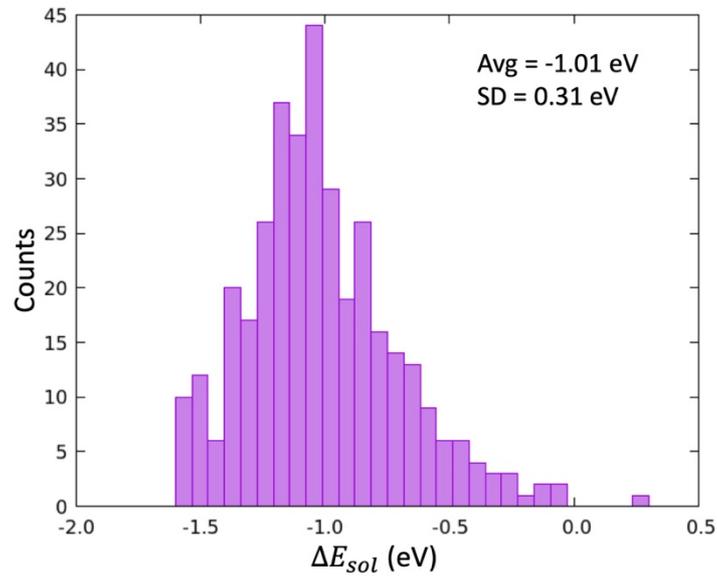

Fig 6. Energy of solution for octahedral sites.

Table 2. Formation energies per carbon atom of the monocarbides, and energy of solution of interstitial carbon in the dilute limit for the pure metals (bcc structured Nb and V, and hcp structured Hf, Ti and Zr) and $M_i$-rich average in the HEA.

|  | Monocarbides (XC) (eV/C atom) | Elemental reference (eV/C atom) | HEA $M_i$-rich average (eV/C atom) |
|---|---|---|---|
| Hf | -2.01 | -1.84 | -1.01 |
| Nb | -1.05 | -0.67 | -0.85 |
| Ti | -1.71 | -1.67 | -1.23 |
| V | -0.93 | -0.74 | -1.03 |
| Zr | -1.83 | -1.73 | -0.94 |

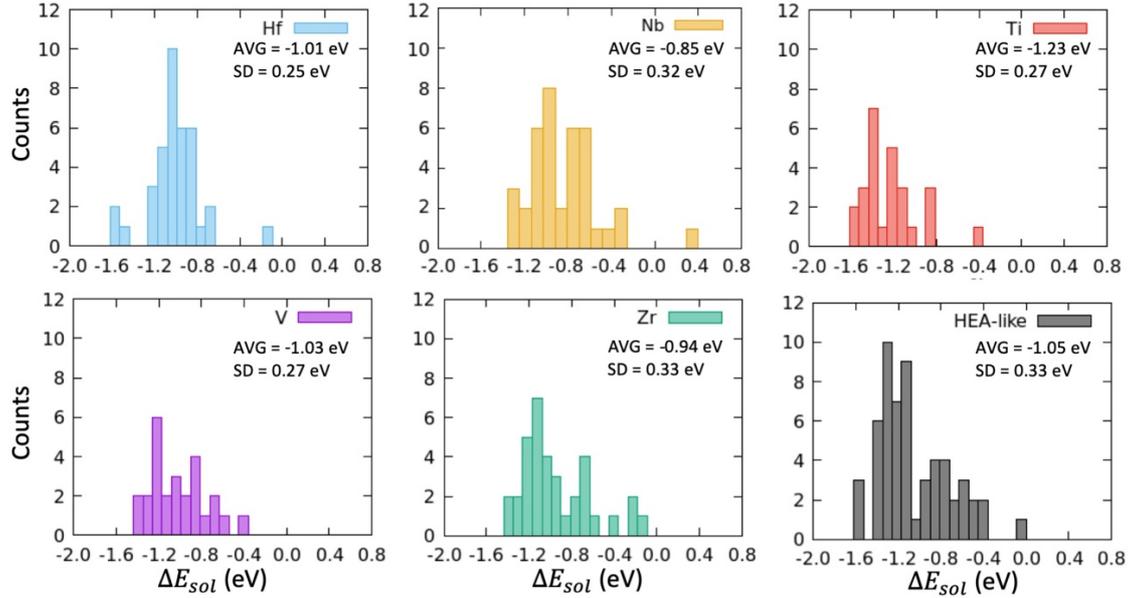

Fig 7. Distribution of the solution energies of carbon atoms in octahedral sites for different $M_i$-rich, as well as the HEA-like local environments.

*3.3 Effect of interstitial carbon on local displacements in HfNbTiVZr HEA*

Figure 8 shows the displacements of the metal atoms in the supercell from their positions in the relaxed carbon free HEA supercell when the carbon interstitial is introduced. In this figure we plot the displacement as a function of their distance to the carbon atom for all octahedral sites. It is possible to observe the different neighbor shells, and in particular the first neighbor shell is clearly distinguishable. In general, the largest displacements are found in the first coordination sphere, with decreasing magnitude as the positions are farther away from the interstitial. None the less, there are still some large displacements for atoms located at 4 Å from the carbon interstitial, and even the atoms located farthest away from the carbon interstitial present considerable displacements.

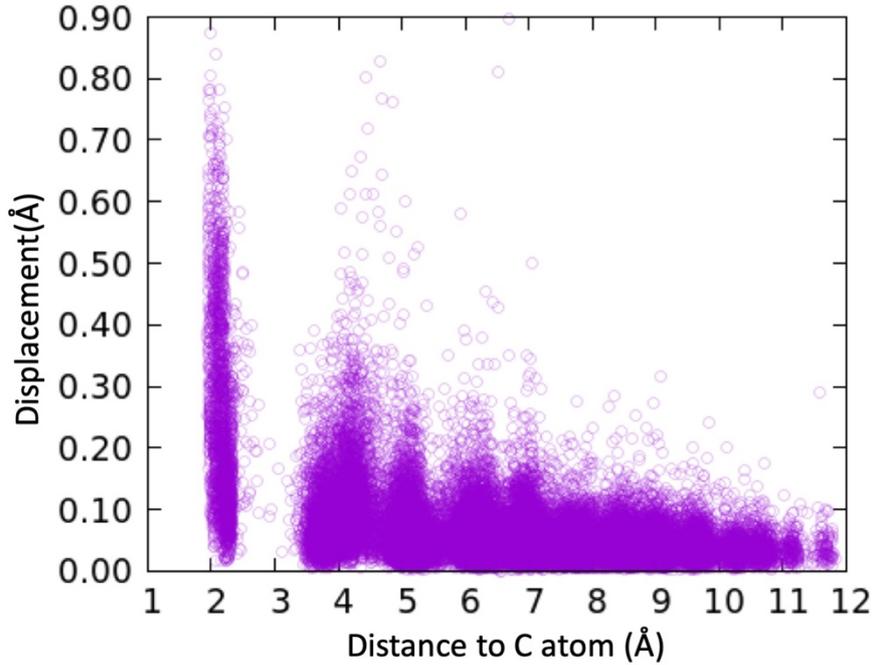

Figure 8. Atomic displacements of metal atoms upon carbon inclusion as a function of the distance to the carbon interstitial.

For comparison, we have obtained the atom displacements in the case of a carbon atom in bcc Nb and V, since they are the two constituent elements that have a stable bcc structure at low temperature. In Figure 9a, the displacements as a function of distance from the carbon atom are compared, and in Figure 9b we show a schematic representation of the displacements of pure Nb, where the displacement vectors have been scaled up for clarity. Figure 9b shows that the largest displacements in the pure elements correspond to the atoms in the short axis of the nearest neighbor octahedron, which are pushed away from the carbon interstitial, while the other 4 nearest neighbors are pulled towards it. In the case of the HEA, it is not possible to discern between these two types of nearest neighbor atoms due to the wide range of nearest neighbor distances. Thus, the average displacement of the first shell in HfNbTiVZr is lower in magnitude and has a large error bar. For the pure element, the displacement tends to become 0 with increasing distance, while for the HEA, even the positions at the greatest distance from C that is possible to investigate with our supercell, have a finite displacement.

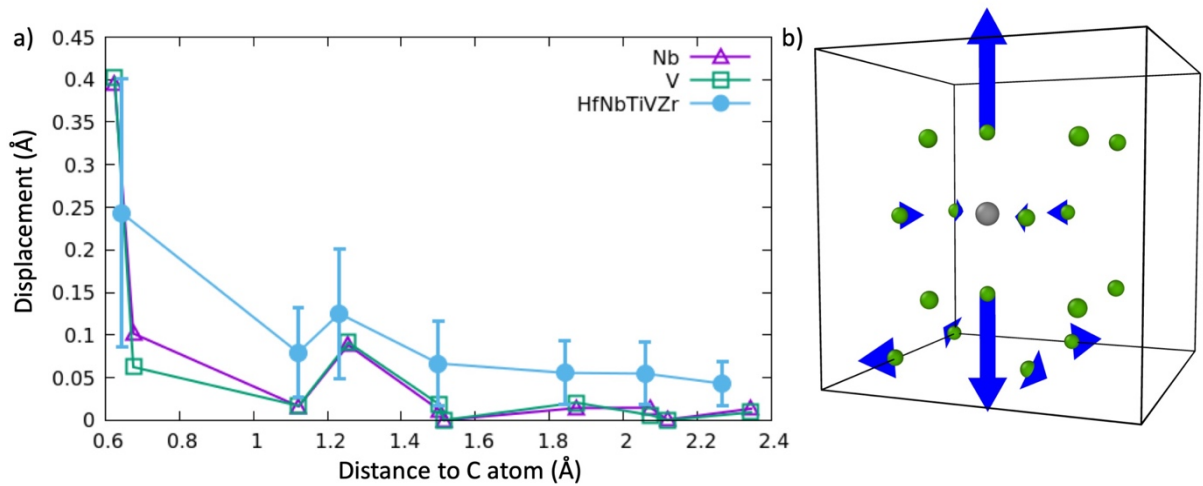

Figure 9. a) Comparison of displacements of metal atoms in pure Nb, V and HfNbTiVZr as a function of distance to the carbon interstitial. b) Schematic illustration of atomic displacements in bcc Nb.

*3.4 Local environment and $\Delta E_{sol}$*

To investigate the effect of local chemical environment on the solution energies, we have studied its correlation with volume and local valence electron concentration (VEC). Figure 10 presents the plotted $\Delta E_{sol}$ versus the nearest neighbor polyhedral volume, the Voronoi volume, the radical Voronoi volume and VEC. The VEC has showed a good correlation with the energy of solution in previous studies [24,41]. The VEC is computed as the average of valence electron numbers over the atoms in the first nearest neighbor shell of an interstitial carbon atom, with 4 valence electrons for Hf, Ti, Zr and 5 for Nb and V. For HfNbTiVZr the similar valence electrons for the constituent elements make the VEC a bad descriptor since it is unable to distinguish the different chemical environments. However, among these properties the Voronoi volume shows the best correlation with the solution energies.

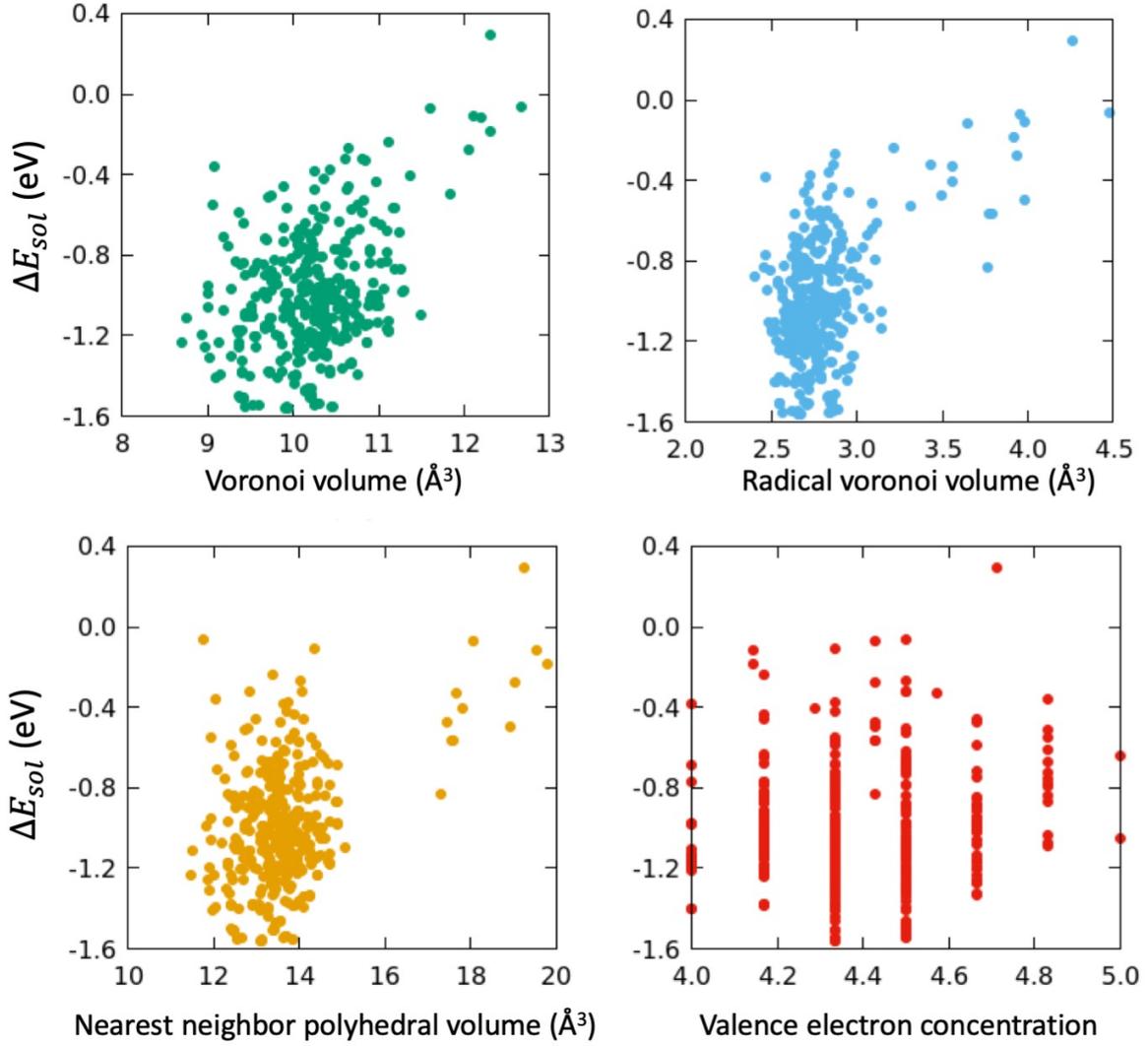

Figure 10. $\Delta E_{sol}$ versus Voronoi volume, radical Voronoi volume, nearest neighbor polyhedral volume and valence electron concentration.

Finally, the local chemical environment of carbon atoms in octahedral sites has been evaluated with a local cluster expansion method [42,43]. We have tried two approaches for the cluster expansion. In the first one, only the on-site interactions of the number of first nearest neighbors of each element type have been used. In the second approach an effective interaction for the Voronoi volume has been added to the first approach. The expressions for these two approaches are given in Equations 2 and 3, respectively. Both approaches fail to describe the individual values of the solution energies. In Figure 11 we have plotted the comparison of the calculated energies using DFT and the predicted energies. The failure of these simplistic approaches is an indication that the energy of solution has a complex dependence on chemical environment. It is needed to account for interactions well beyond the first neighbor shell and possibly the particular configurations of metals around carbon, as was found for Ti and Al around N in reference [44]. This finding is in line with the finding of long-range induced structural relaxations in figures 8 and 9.

$$E_{Sol} \cong \varepsilon_{Hf} n_{Hf} + \varepsilon_{Nb} n_{Nb} + \varepsilon_{Ti} n_{Ti} + \varepsilon_V n_V + \varepsilon_{Zr} n_{Zr} \quad (2)$$

$$E_{Sol} \cong \varepsilon_{Hf} n_{Hf} + \varepsilon_{Nb} n_{Nb} + \varepsilon_{Ti} n_{Ti} + \varepsilon_V n_V + \varepsilon_{Zr} n_{Zr} + \varepsilon_{Vol} V_i \quad (3)$$

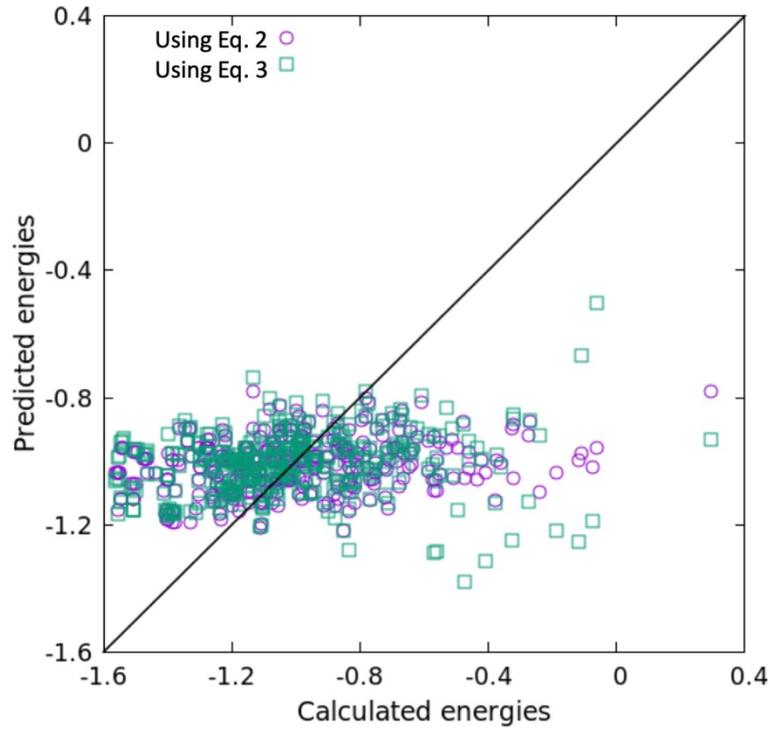

Figure 11. Calculated versus predicted energies using on-site interactions obtained by a local cluster expansion.

4. Conclusions

In this work we have calculated from first principles the energy and structural effects of solution of dilute carbon interstitials in HfNbTiVZr HEA and assessed the impact of local chemical environment on these energies by probing 720 tetrahedral sites and 360 octahedral sites. An excellent agreement between the pair distribution function from the SQS supercell and that from experimental neutron total scattering results has been obtained. We found that carbon atoms initially placed in tetrahedral positions are unstable and move to octahedral or other interstitial positions after relaxation. The solution energies of C in HfNbTiVZr with respect of HEA and graphite display a broad distribution, with an average value of -1.01 eV and a standard deviation 0.31 eV. Both, the average value and minimum value of -1.56 eV are significantly lower than the calculated solution energies of C in pure bcc Nb and V. The energies of solution were negative for all octahedral sites except one with Nb rich local environment. Among all environments that are rich in a single metal type, Ti-rich presents the most favorable sites for carbon interstitials. The introduction of carbon interstitials produces large distortions in the system that extend beyond the first and second nearest neighbors shells. The wide range of displacements found in the nearest neighbor shell, and the failure of the simple local cluster expansion approaches indicate that interaction with atoms beyond the first nearest neighbor shell are needed to understand the complex distribution of local solution energetics for this material system.

**Acknowledgements**

This work was supported by the Swedish Research Council (VR) project No. 2018-04834. Support from the Swedish Government Strategic Research Area in Materials Science on Functional Materials at Linköping University, Faculty Grant SFOMatLiU No. 2009 00971 and from the Knut and Alice Wallenberg Foundation (Wallenberg Scholar Grant No. KAW-2018.0194) is gratefully acknowledged. B.A. acknowledges financial support from the Swedish Foundation for Strategic Research through the Future Research Leaders 6 program, FFL 15-0290 and from the Swedish Research Council (VR) through the grant 2019-05403, Theoretical analysis of the results of electronic structure calculations was funded by RFBR, project number 20-02-00178. Calculations were performed using supercomputer resources provided by the Swedish National Infrastructure for Computing (SNIC) at the National Supercomputer Center (NSC). M.M.N., M.H.S. and B.C.H. acknowledge funding by the NordForsk Nordic Neutron Science Programme through the functional hydrides (FunHy) project (grant number 81942).